\documentclass[journal]{IEEEtran}
\IEEEoverridecommandlockouts
\usepackage{amsmath,amssymb,amsfonts}
\usepackage{algorithmic}
\usepackage{graphicx}
\usepackage{textcomp}
\usepackage{xcolor}
\usepackage{comment}
\usepackage{multicol}
\usepackage{xurl}
\usepackage{multirow}
\usepackage{color}
\usepackage{subcaption}
\usepackage{graphicx}  
\usepackage{soul}
\usepackage{physics}
\usepackage{booktabs}

\usepackage[ruled,vlined,linesnumbered]{algorithm2e}
\usepackage{comment}
\usepackage[numbers,sort&compress,square]{natbib}
\usepackage{enumerate}
\usepackage{titlesec}
\titlespacing*{\section}
{0pt}{1 mm plus 1mm minus 1mm}{1 mm plus 1mm minus 1mm}
\titlespacing*{\subsection}
{0pt}{1 mm plus 1mm minus 1mm}{1 mm plus 1mm minus 1mm}

\usepackage[font=footnotesize]{caption}
\usepackage[font=footnotesize]{subcaption}
\usepackage[numbers,sort&compress,square]{natbib}
\usepackage{pdfpages}

\DeclareCaptionFormat{mysubcaption}{\footnotesize#1#2#3\par}
\usepackage[bookmarks=false, hidelinks]{hyperref}
\usepackage[normalem]{ulem}
\usepackage{float}
\usepackage{scalerel}
\usepackage{tikz}
\usetikzlibrary{svg.path}

\definecolor{orcidlogocol}{HTML}{A6CE39}
\tikzset{
  orcidlogo/.pic={
    \fill[orcidlogocol] svg{M256,128c0,70.7-57.3,128-128,128C57.3,256,0,198.7,0,128C0,57.3,57.3,0,128,0C198.7,0,256,57.3,256,128z};
    \fill[white] svg{M86.3,186.2H70.9V79.1h15.4v48.4V186.2z}
                 svg{M108.9,79.1h41.6c39.6,0,57,28.3,57,53.6c0,27.5-21.5,53.6-56.8,53.6h-41.8V79.1z M124.3,172.4h24.5c34.9,0,42.9-26.5,42.9-39.7c0-21.5-13.7-39.7-43.7-39.7h-23.7V172.4z}
                 svg{M88.7,56.8c0,5.5-4.5,10.1-10.1,10.1c-5.6,0-10.1-4.6-10.1-10.1c0-5.6,4.5-10.1,10.1-10.1C84.2,46.7,88.7,51.3,88.7,56.8z};
  }
}

\newcommand\orcidicon[1]{\href{https://orcid.org/#1}{\mbox{\scalerel*{
\begin{tikzpicture}[yscale=-1,transform shape]
\pic{orcidlogo};
\end{tikzpicture}
}{|}}}}

\def\BibTeX{{\rm B\kern-.05em{\sc i\kern-.025em b}\kern-.08em
    T\kern-.1667em\lower.7ex\hbox{E}\kern-.125emX}}
\begin{document}

\newenvironment{ldescription}[1]
  {\begin{list}{}%
   {\renewcommand\makelabel[1]{##1\hfill}%
   \setlength\itemsep{0pt}%
   \settowidth\labelwidth{\makelabel{#1}}%
   \setlength\leftmargin{\labelwidth}
   \addtolength\leftmargin{\labelsep}}}
  {\end{list}}

\title{Coordinated vs.\ Sequential Transmission Planning}

\author{Maya Domeshek, Christoph Graf \vspace{-10mm}
}
\author{Maya Domeshek, Christoph Graf, Bur\c{c}in \"Unel
\thanks{M.\ Domeshek, C.\ Graf, and B.\ \"Unel are with the Institute for Policy Integrity, New York University School of Law, New York, NY 10012 USA 
(e-mail: md6630@nyu.edu, christoph.graf@nyu.edu, burcin.unel@nyu.edu)}
}

\maketitle

\maketitle
\begin{abstract}
Coordinated planning of generation, storage, and transmission more accurately captures the interactions among these three capacity types necessary to meet electricity demand, at least in theory. However, in practice, U.S.\ system operators typically follow a sequential planning approach: They first determine future generation and storage additions based on an assumed unconstrained (`copper plate') system. Next, they perform dispatch simulations of this projected generation and storage capacity mix on the existing transmission grid to identify transmission constraint violations. These violations indicate the need for transmission upgrades. We describe a multistage, multi-locational planning model that co-optimizes generation, storage, and transmission investments. The model respects reliability constraints as well as state energy and climate policies. We test the two planning approaches using a current stakeholder-informed 20-zone model of the PJM region, developed for the current FERC Order No.~1920 compliance filing process. In the model specification that most closely matches PJM's current data inputs, we find that the co-optimized approach estimates 84\% lower transmission upgrade needs than the sequential model, leading to total system costs that are 3.7\% lower while having no impact on reliability and similar greenhouse gas emissions. Our sensitivities uniformly show transmission and total cost savings from co-optimized planning and similar reliability and climate outcomes. 
\end{abstract}


\section{Introduction}
\IEEEPARstart{T}{he} U.S. Federal Energy Regulatory Commission (FERC) tasks ISOs with creating regional transmission plans. Historically, under FERC Order No.~1000, these have been short-term transmission plans on the order of a 10-year time horizon. But more recently, FERC Order No.~1920 has required ISOs to engage in long-term planning on a 20-year time horizon that takes into account policies affecting supply, demand, and decarbonization. There is substantial variation in how ISOs plan transmission. But common practice among U.S.\ ISOs is to plan future transmission systems in a sequential manner \cite{Pjm25}. This planning process consists of two steps: First, planners project future generation and storage capacity, either informed by the interconnection queue or by using a resource capacity expansion model that incorporates market trends and (state) policies, typically under the assumption of unlimited transmission capacity (`copper plate'). When using a copper plate capacity expansion model, aggregate generation and storage capacity must then be spatially distributed. Second, planners conduct power flow analyses using the existing transmission network and the projected generation and storage mix from the first step. Any violations of transmission constraints identified in this step signal the need for transmission upgrades.

The sequential approach can work reasonably well for short- to near-term transmission upgrades, as planners typically have good visibility into the timing and location of resources expected to come online in the next few years. However, for longer-term planning, such as the 20-year horizon required by FERC Order No.~1920, this approach becomes impractical. This shortcoming is especially true in the context of decarbonization, which requires transformative changes in the generation mix. In such cases, interconnection queues are often filled with speculative or `zombie' projects \cite{UK25}, where developers are simply testing the waters to assess potential costs of interconnection at different locations. As a result, the interconnection queue may provide only a weak signal of actual future generation and storage entry, making it an unreliable foundation for long-term transmission planning. At the same time, because decarbonization will require new renewable generation that may be located far from load, resource expansion modeling that ignores transmission may not locate the new generators in the most efficient way. 

A recent complaint filed with FERC by five public service commissions regarding the practices of another U.S. system operator---MISO's `Long Range Transmission Planning,' Tranche 2.1 \cite{miso25}, which involves approximately \$22 billion in transmission investments \cite{miso_tranche21_24}---has brought greater attention to the details of transmission planning, including the potential shortcomings of sequential planning. We contribute to this discussion by quantifying the benefits of integrated, co-optimized planning for resources (generation and storage) and transmission, compared to sequential planning using PJM as a case study.

Integrated co-optimized planning for resources (generation and storage) and transmission accounts for the interdependency between resources and transmission \cite{KrHo15, Shu2025}. For instance, generation can be located closer to load centers or sited farther away if paired with adequate transmission infrastructure. It is the combination of transmission and generation investment and operating costs, along with system, reliability, and policy constraints, that determines the most cost-effective outcome. In the academic literature, co-optimization models have long been used as transmission planning tools. For example, \cite{SpHo17} shows the benefits of a co-optimized transmission planning model for a 24-bus representation of the U.S.\ Eastern Interconnections relative to a reactive sequential transmission planning model. Consistent with \cite{MaLi19}, who show the benefits of co-optimization in the U.S.\ Pacific Northwest, both studies find significant net present value benefits when co-optimizing. Co-optimized planning can also be used as a benchmark against which to measure the performance of non-co-optimized systems. For example, \cite{Senga2026} measure the cost and reliability performance of federal legislation requiring inter-regional transmission with endogenously chosen levels of interregional transmission and generation. An alternative to integrated co-optimized planning is proactive or anticipatory transmission planning, where the transmission planner anticipates how generation investment and operations will respond to changes in the transmission network \cite{SaOr06,GaCo09,SpHo17}. This framework uses bi-level programming and allows the study of more general cases, including those where generators may possess market power \cite{SaOr06,WoTe21}. However, this generalization comes at the cost of increased computational complexity. Under the assumptions of perfect competition, welfare maximization, and efficient transmission pricing, the proactive planning approach theoretically yields results equivalent to those of integrated planning \cite{GaCo09,vdW12,SpHo17}.  

There are two bookends for modeling baseline capacity and storage additions across a region. At one end, resource capacity additions can be modeled assuming an unconstrained regional grid, a `copper plate' approach. At the other, they can be modeled using existing transmission corridor limits. The former approach prioritizes regional resource selection and may lead to overinvestment in transmission, while the latter typically favors local siting and may understate transmission needs. For example, \cite{SpHo17} compares co-optimizing resource and transmission expansion with a sequential approach where generation is added based on the existing transmission network. They find that co-optimization results in greater transmission investment, as it more accurately reflects the interactions between generation and transmission. However, in practice, most U.S. ISOs, such as PJM and MISO \cite{Pjm25}, typically use the unconstrained grid assumption when planning transmission, which can lead to an overestimation of transmission requirements.

Modeling approaches in the academic literature comparing integrated or co-optimized resource and transmission expansion solutions to solutions that are sequentially derived typically start planning resources using the existing configuration of the transmission system \cite{SpHo17, MaLi19}. We start with a resource plan that is based on an assumed unconstrained network consistent with how U.S.\ ISOs are currently planning transmission.

We built a multi-stage, multi-zonal, generation, storage, and transmission planning model (GS\&TEP) similar to \cite{munoz2013engineering,qiu2016stochastic,rafaj2007internalisation,chen2018advances,rodgers2019assessing,quiroga2019power,chen2019multi,chiu2020future,lv2020generation,pereira2020power,gbadamosi2020multi,fitiwi2020enhanced,sani2021decarbonization,verastegui2021optimization, KhGr24} using PJM as a case study. However, unlike existing studies, our approach places particular emphasis on modeling state policies, including both local and ISO-wide renewable portfolio standards, whether technology-neutral or resource-specific (carve-outs), as well as local resource specific targets for renewable energy and storage capacity. This is important, as decarbonization of the grid due to state policies is one of the reasons that sequential planning may no-longer be adequate. Our case study is built on a snapshot from the stakeholder-derived PJM model used in the current FERC Order No.~1920 compliance filing process \cite{Pjm25}. The geographical resolution is consistent with PJM's load zones covering 20 zones. We model all 8760 consecutive hours within each modeled year for a given weather year and incorporate a resource adequacy constraint to ensure reliability beyond that specific year. This constraint aligns with PJM’s planning standards by using the effective load carrying capability (ELCC) at the technology level to convert available nameplate capacity into reliable capacity, which in aggregate needs to meet a pre-specified reliability target. 

While this is not the first paper to look at the benefits of co-optimized planning, it is the first to do so in PJM with attention to the institutional details of the region, including state policies and the capacity market. It also closely follows PJM's own data curation for the Order No.~1920 transmission planning process where a need for co-optimized planning may be particularly acute. We find that co-optimized resource and transmission planning yields net present value benefits around \$23 billion~(2024\$) over twenty years. Co-optimized planning also yields reduced transmission need and generally builds resources closer to load. We find that co-optimized solutions lead to similar reliability and climate outcomes as sequential solutions. 

The remainder of this paper is organized as follows. Section \ref{sec:case_study} describes PJM. Section \ref{sec:setup} gives a brief description of the model, scenario setup, and data. A more detailed model description is available in the Appendix \ref{appendix:ap}. Section \ref{sec:r&d} includes results and discussion. Finally, Section \ref{sec:conclusion} concludes the paper.

\section{PJM Case Study} \label{sec:case_study}
PJM is the Independent System Operator covering all or part of 13 states (DE, MD, NJ, OH, PA, VA, WV and parts of IL, IN, KY, MI, NC, TN) and the District of Columbia (DC). Some of the states in the region have restructured their electricity sectors while others maintain vertically integrated utilities. PJM coordinates both the regional wholesale market and a regional capacity market to ensure long-term reliability. The region has a notoriously long interconnection queue. In recent years, it has also had strong load growth, primarily from data centers. It has states with climate policies promoting the transition from fossil fuels to clean resources (DC, DE, MD, NJ, PA, VA, IL, MI, NC). It also has states without climate policies (OH, WV, IN, KY, TN). These institutional features make PJM a particularly rich setting in which to study transmission planning.

In the past, PJM has created its transmission plan through the FERC Order No.~1000 Regional Transmission Expansion Plan (RTEP) process in which it uses information about generation in the interconnection queue and integrated resource plans from vertically integrated utilities to develop a short-term generation and storage capacity projection and build a transmission grid that complements that future resource capacity mix. It is currently developing its plan for FERC Order No.~1920, which requires the creation of long-term planning scenarios incorporating state policies affecting supply, demand, and decarbonization as well as economic and technological trends over 20 years. PJM is proposing a sequential modeling framework under which it first runs a generation and storage capacity expansion model, then assigns projected capacity to locations on the grid and looks for transmission violations \cite{Pjm25}. 

\section{Modeling and Data} \label{sec:setup}

\subsection{Model}
For this paper, we developed a Generation, Storage, and Transmission Expansion Model (GS\&TEP) described in Appendix \ref{appendix:ap}. We run two main types of scenarios, ``co-optimized'' scenarios and ``sequential'' scenarios. For the co-optimized scenarios, we run the standard version of the model with both transmission and resource capacity expansion (GS\&TEP) (see~Appendix \ref{appendix:ap}.\ref{subsection:coopt_model}). For the sequential planning scenarios, we first run the Generation and Storage Expansion Model (GSEP) on a `copper plate' which plans capacity expansion without regard to transmission needs (see~Appendix \ref{appendix:ap}.\ref{subsection:copper_plate}). Because some state policies require renewable capacity to be build within state boundaries, investment costs vary geographically, and wind and solar resource availability as well as load factors differ by location, the copper-plate formulation captures these spatial heterogeneities in costs, policies, and resource quality but does not endogenously determine transmission expansion or congestion between locations. We then run the standard version of the model, but fixing the resource capacity additions from the GSEP model, making it only a Transmission Expansion (TEP) model. 

The model includes constraints that provide a detailed representation of state-level climate policies (see~Appendix \ref{appendix:ap}.\ref{subsubsection:pol_const}). A description of the state-level policies can be found in Section~\ref{subsubsection:policy_assumptions}.

The model can be run with two different reliability constraints: the `ELCC' constraint and the `Rsv Req' constraint. The `ELCC' constraint (Appendix~\ref{appendix:ap}.\ref{subsection:reliability}, Equation \ref{equation:cap_market}) mimics PJM's capacity market by requiring the model to build enough reliable capacity to meet 115\% of peak demand where reliable capacity means the name plate capacity multiplied by a technology and year specific Effective Load Carrying Capacity (ELCC) determined by PJM. This constraint tends to push the model to build more generation capacity, and especially gas generation capacity. The `Rsv Req' constraint (Appendix~\ref{appendix:ap}.\ref{subsubsection:power_balance}, Equation \ref{equation:sup_dem_eq} or Equation \ref{equation:sup_dem_reg}) requires the model to generate 115\% of load in each hour . This constraint allows the model to use more renewables and transmission to meet reliability because those resources can contribute to reliability in any hour they are available. 

We ran the co-optimized and sequential versions of the model with the two different reliability constraints and with high and low electricity demand and high and low gas prices, yielding a total of sixteen scenarios described in the sensitivity analysis section \ref{subsubsection:sensitivities}. The bulk of the results section focuses on the `ELCC + High Gas Price + High Demand' case as the one that most closely resembles PJM's own modeling. 

\subsection{Data Setup}
This case study relies on public data from NREL ReEDS \cite{Brown_Regional_Energy_Deployment} and on data provided by Energy Exemplar to PJM through its Clean Attribute Procurement Senior Taskforce process \cite{CAPSTF}. It also includes some assumptions taken from PJM's Long Term Regional Transmission Planning Process \cite{pjm10Dec2024TEAC} and the scenario assumptions for PJM's FERC Order No.~1920 cases \cite{pjm10Apr2025TEAC, pjm5Sept2025TEAC}. As discussed above, the model consists of 20 zones within the PJM footprint. Each zone has an initial set of generator and storage capacity with some anticipated retirements as well as initial capacity of transmission corridor limits. To optimize over the 20 year time horizon, the model requires information about the cost to operate existing plants, the cost to build and operate new plants, hourly availability or capacity factors for each technology, the cost of increasing transmission corridor limits, hourly location-specific load factors, projected location-specific peak demand levels, reliability contributions of resources along with reliability targets, and state policy designs and targets. 

\subsubsection{Technology Characteristics and Investment Options}
Initial capacity of generation and storage in each zone is taken from \cite{CAPSTF}, as are the plant characteristics (heat rates, variable operations and maintenance (O\&M) costs, fixed O\&M costs). Fuel price projections are taken from NREL ReEDS \cite{Brown_Regional_Energy_Deployment}, and when multiplied by heat rates, provide the fuel costs of fossil plants.

The model allows investment in solar, onshore wind, offshore wind, new gas combined cycle, new gas combustion turbines, and four-hour batteries. Overnight capital costs, variable and fixed O\&M and heat rates for these technologies are taken from NREL ATB 2024 \cite{NRELATB2024}. Capital costs are zone-specific, based on scaling factors provided by PJM in \cite{pjm10Apr2025TEAC}. Capital recovery factors and capital recovery periods are from \cite{pjm10Dec2024TEAC}. We set limits on solar and onshore wind capacity described in \cite{pjm5Sept2025TEAC}. The model also allows for endogenous retirement of existing generation.

Solar and wind availability vary across hours and zones, impacting the cost-effectiveness of investing in capacity in different zones. These availability factors are based on historical data, in this case NREL's availability factor data for historical year 2008 \cite{Brown_Regional_Energy_Deployment}. As with existing gas plants, new gas plants have fuel costs dependent on their heat rates and the NREL ReEDS gas price projections \cite{Brown_Regional_Energy_Deployment}. Batteries are assumed to have an 85\% round-trip efficiency.

\begin{figure}
    \includegraphics[width=0.475\textwidth]{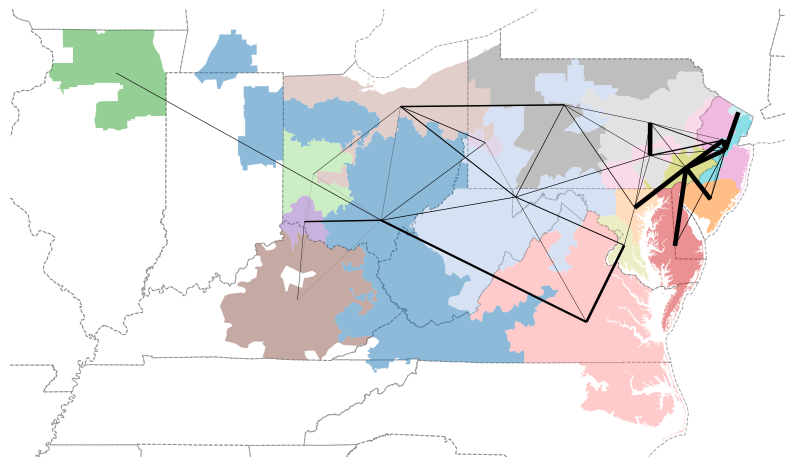}
    \caption{Baseline Topology and Transmission Corridor Limits}
    \label{fig:trans_map_bl}
\end{figure}

The initial capacity limit of transmission corridors between zones correspond to those in \cite{CAPSTF}. Fig.~\ref{fig:trans_map_bl} shows the baseline topology and transmission corridor limits. Transmission expansion is assumed to occur through corridor-reinforcement (reconductoring) at a cost of \$2076/MW-mile (2024\$) \cite{ho2021regional} which may be lower-cost than transmission expansion costs assumed by PJM.

\subsubsection{Load}
The load factor (which is the share of peak load in each zone and hour for the 8760 hours of the year) is taken from historical load years and is the same across all epochs. For these model runs, we use historical year 2008 from NREL ReEDS \cite{Brown_Regional_Energy_Deployment}. NREL provides historical load information for the years 2007--2013 and 2016--2024. We have run the model using load factors based on each of those years, but we chose the 2008 year because it produces the model runs with the highest costs, indicating that it is the most difficult set of load factors for the model to meet. Total load for each epoch is determined by multiplying the load factors by PJM's peak load projections. The model currently uses PJM's peak load from PJM's `2025 Long-Term Load Forecast Report' \cite{PjmLoad25} as the high demand scenario and PJM's peak load from PJM's `2022 Long-Term Load Forecast Report' \cite{PjmLoad22} as the low demand scenario. 

\subsubsection{Policy Assumptions} \label{subsubsection:policy_assumptions}
We originally took state RPS levels from \cite{CAPSTF} but updated to the most recent available information as of April 2025 (Table\;\ref{rps_table}). Most RPS policies are represented as regional requirements, where the regional requirement is the demand weighted sum of all state-level requirements. Two states, IL and VA, which have in-state RPS requirements written into law are subject to a special RPS constraint that require a portion of RPS generation to occur in-state (92\% and 75\% respectively). State renewable capacity targets (Table\;\ref{cap_tgt_table}) were also originally taken from \cite{CAPSTF} but updated with information from \cite{pjm10Apr2025TEAC}.

\begin{table}
    \begin{center}
    \caption{State RPS Policies [\% of Load]}
    \label{rps_table}
    \begin{tabular}{ |c|c|c|c|c|c| } 
    \hline
    State                       & Technology          & 2030 & 2035 & 2040 & 2045  \\
    \hline
    \hline
    \multirow{4}{3em}{DC}        & Solar, Wind, Hydro &      &      &      & \\ 
                                 & Solar, Wind        &87    &100   &100   &100 \\
                                 & Solar              &5     &10    &14    &15 \\
                                 & Wind               &      &      &      & \\
    \hline
    \multirow{4}{3em}{DE}        & Solar, Wind, Hydro &28   &40     &40    &40 \\ 
                                 & Solar, Wind        &     &       &      & \\
                                 & Solar              &5    &10     &10    &10 \\
                                 & Wind               &     &       &      & \\
    \hline
    \multirow{4}{3em}{IL}        & Solar, Wind, Hydro &34   &38     &43    &43 \\ 
                                 & Solar, Wind        &     &       &      & \\
                                 & Solar              &19   &21     &23    &23 \\
                                 & Wind               &15   &17     &19    &19 \\    
    \hline
    \multirow{4}{3em}{MD}        & Solar, Wind, Hydro &50   &50     &50    &50 \\ 
                                 & Solar, Wind        &     &       &      & \\
                                 & Solar              &15   &15     &15    &15 \\
                                 & Wind               &     &       &      & \\
    \hline
    \multirow{4}{3em}{NJ}        & Solar, Wind, Hydro &     &       &      & \\ 
                                 & Solar, Wind        &50   &50     &50    &50 \\
                                 & Solar              &2    &       &      & \\
                                 & Wind               &     &       &      & \\
    \hline
    \multirow{4}{3em}{PA}        & Solar, Wind, Hydro &     &       &      & \\ 
                                 & Solar, Wind        &8    &8      &8     &8 \\
                                 & Solar              &1    &1      &1     &1 \\
                                 & Wind               &     &       &      & \\
    \hline
    \multirow{4}{3em}{VA}        & Solar, Wind, Hydro &38   &59     &79    &100 \\ 
                                 & Solar, Wind        &     &       &      & \\
                                 & Solar              &     &       &      & \\
                                 & Wind               &     &       &      & \\
    \hline
    \end{tabular}
    \end{center}
\end{table}

\begin{table}
    \begin{center}
    \caption{Cumulative State Renewable Capacity Targets [MW]}
    \label{cap_tgt_table}
    \begin{tabular}{ |c|c|c|c|c|c| } 
    \hline
    State   & Technology    & 2030 & 2035 & 2040 & 2045  \\
    \hline
    \hline
            & Storage       &   &  & & \\ 
    DE      & Offshore Wind &   &1,200  &1,200 &1,200\\
            & Solar         &   &  & & \\
    \hline 
            & Storage       & 1,500 &3,000&3,000 &3,000\\ 
    MD      & Offshore Wind &   & 8,500 &8,500 &8,500\\
            & Solar         &   &  & & \\
    \hline
            & Storage       &2,500   &2,500  &2,500&2,500\\ 
    MI      & Offshore Wind &   &  & & \\
            & Solar         &   &  & & \\
    \hline
            & Storage       &2,000   &2,000  &2,000  &2,000\\ 
    NJ      & Offshore Wind &2,758   &10,100 &11,200 & 11,200\\
            & Solar         &5,230   &5,230  &5,230  &5,230\\
    \hline
            & Storage       &1,350   &3,100  &3,100 &3,100\\ 
    VA      & Offshore Wind &        &5,200  &5,200 &5,200\\
            & Solar         &        &  & & \\
    \hline

    \hline
    \end{tabular}
    \end{center}
\end{table}
    
We assume coal plants will persist after 2040 despite EPA's 2024 111(d) rule, because of the federal government's recent attempt to rescind that rule, except in IL where state-law requires their retirement by 2030. IL state law also requires the progressive retirement of all existing gas capacity so we retire 80\% of IL gas by 2035, 90\% by 2040 and 100\% by 2045 with all gas built after 2025 getting a shortened payback period.

\subsubsection{Other Model Parameters and Specifications}
The time discount rate for the model is set at 7.2\% based on \cite{pjm10Apr2025TEAC}.

For scenarios with capacity market reliability constraints, we took ELCC ratings for different technologies from PJM's ELCC ratings for Delivery Year 2026/27-2034/35 and held constant thereafter \cite{ELCCratings}.

\subsubsection{Simplifications and Limitations}
We make several assumptions to keep the model computationally tractable. We use a 20-zone representation of the PJM system and model transmission corridors, rather than specific lines. These publicly available data may be considerably inferior to the data accessible by transmission planners. The comparison between sequential planning and co-optimized planning does not take into account that the sequential models used by system operators typically run detailed power flow studies to identify transmission upgrade needs \cite{KhGr24}.

Our model assumes perfect foresight of demand growth and available generation and storage technologies and their investment costs. Furthermore, our model represents new generation decisions as aggregate capacity within a given area, rather than a standalone unit, which limits the ability to model operational constraints such as ramping or unit-commitment.

Unlike \cite{KhGr24}, we do not explicitly optimize the offshore wind grid; instead, we assume that offshore wind generation will be integrated into each zone.

We use annual technology-level ELCCs to ensure our results align with a specific reliability target. In practice, however, these ELCCs are high-dimensional functions that depend not only on the capacity of each technology but also on the capacities of other technologies including transmission upgrades.

\section{Results and Discussions} \label{sec:r&d}

We find that co-optimized planning leads to lower transmission build and lower system costs. Across our sensitivities, co-optimized scenarios either have no impact on reliability or improve reliability while having emissions outcomes similar to sequential cases.

\subsection{Impact on Transmission and Capacity Build} \label{subsubsection:trans_cap_res}
\begin{figure}[h]
    \begin{subfigure}[h]{.475\linewidth}
        \includegraphics[width=\textwidth]{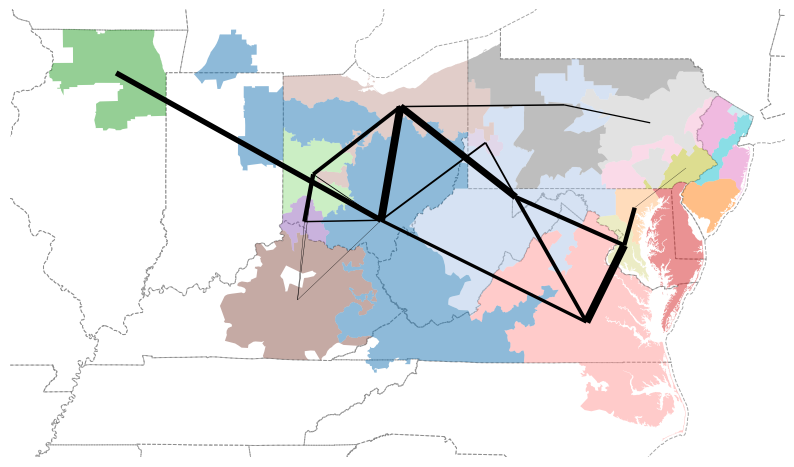}
        \caption{Sequential}
        \label{fig:trans_map_seq}
    \end{subfigure}%
    \hfill
    \begin{subfigure}[h]{.475\linewidth}
        \includegraphics[width=\textwidth]{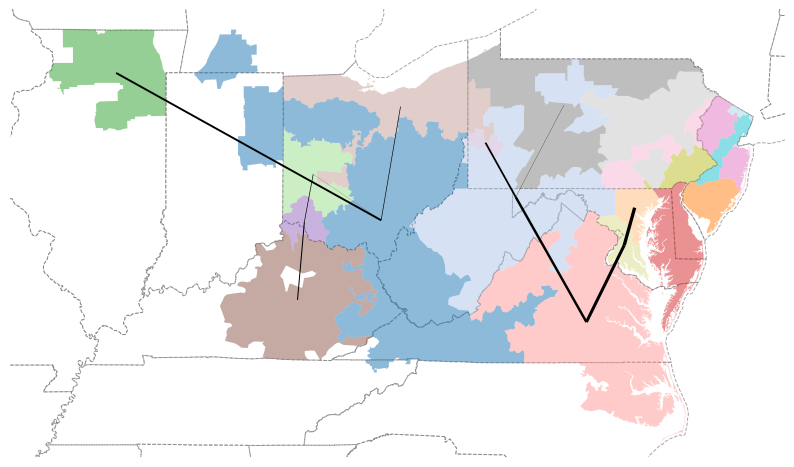}
        \caption{Co-optimized}
        \label{fig:trans_map_coopt}
    \end{subfigure}
    \caption{Cumulative Added Transmission 2045 [GW]}
    \label{fig:trans_map}
\end{figure}

In Fig.\;\ref{fig:trans_map_coopt}, we show the co-optimized transmission capacity expansion needs for the `ELCC + High Gas Price + High Demand' case. Our findings suggest that the optimal planning solution involves reinforcing the transmission corridor between PJM West and Illinois, as well as better integrating the Dominion service territory (primarily VA) to the North (North-South backbone). While similar solutions appear in the sequential planning model (Fig.\;\ref{fig:trans_map_seq}) they tend to be more extensive. The sequential model also identifies additional transmission corridor upgrade needs. In Table\;\ref{tabular:trans_table}, `ELCC + High Gas Price + High Demand,' we show that, by the end of the planning horizon, the sequential model plans for approximately four and a half times the transmission capacity of the co-optimized model.

\begin{figure}
    \begin{subfigure}{.475\linewidth}
      \includegraphics[width=\linewidth]{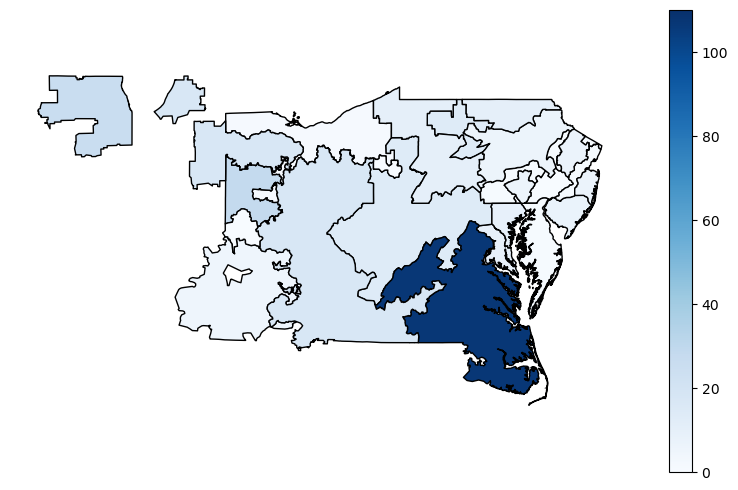}
      \caption{Sequential Renewable}
      \label{re_cap_seq}
    \end{subfigure}
    \hfill 
    \begin{subfigure}{.475\linewidth}
      \includegraphics[width=\linewidth]{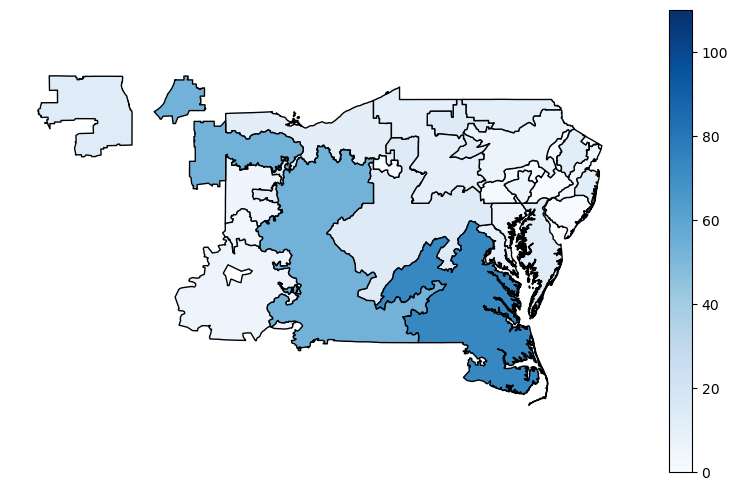}
      \caption{Co-optimized Renewable}
      \label{re_cap_coopt}
    \end{subfigure}
    \medskip 
    \begin{subfigure}{.475\linewidth}
      \includegraphics[width=\linewidth]{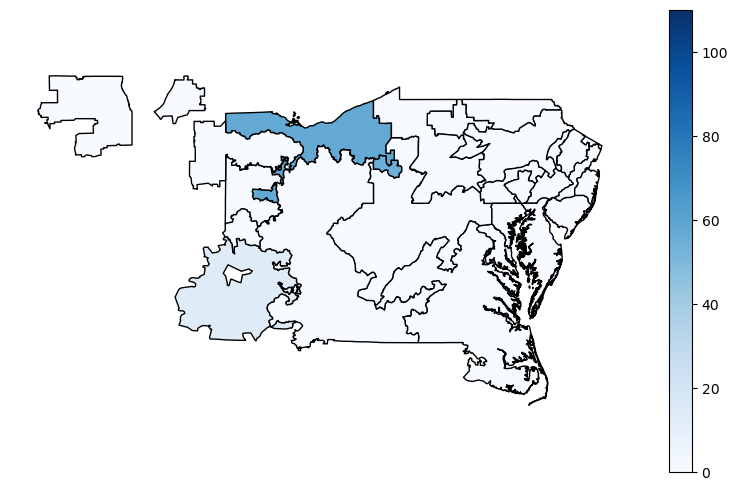}
      \caption{Sequential Gas}
      \label{gas_cap_seq}
    \end{subfigure}
    \hfill
    \begin{subfigure}{.475\linewidth}
      \includegraphics[width=\linewidth]{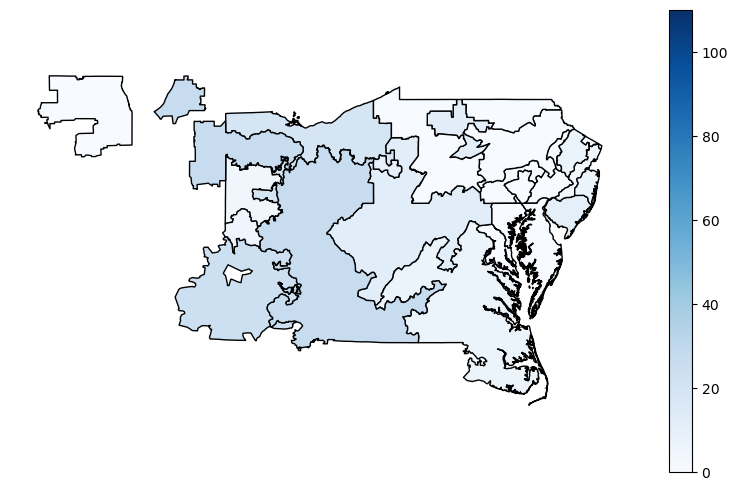}
      \caption{Co-optimized Gas}
      \label{gas_cap_coopt}
    \end{subfigure}
    \caption{Cumulative Renewable and Gas Capacity Additions by 2045 [GW]}    
    \label{cap_map}
\end{figure}

Co-optimization reduces transmission need by spreading added generation capacity more widely around PJM's footprint. Total added gas and renewable capacity is similar for the Sequential and Co-optimized cases. But as shown in Fig.\;\ref{cap_map}, the sequential case concentrates renewables in the southeast of PJM where solar resource availability is most favorable (Fig.\;\ref{re_cap_seq}) and gas in the northwest of PJM where gas capacity is cheapest (Fig.\;\ref{gas_cap_seq}). The co-optimized case builds fewer renewables in the southeast and more in central PJM (Fig.\;\ref{re_cap_coopt}) and spreads gas capacity throughout western PJM (Fig.\;\ref{gas_cap_coopt}). 

\begin{figure}[h]

    \begin{subfigure}[h]{.475\linewidth}
        \includegraphics[width=\textwidth]{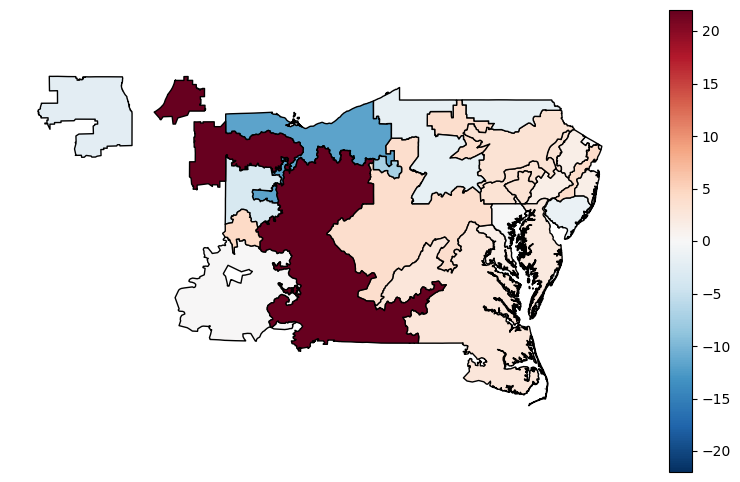}
        \caption{Sequential}
        \label{fig:net_load_seq}
    \end{subfigure}%
    \hfill
    \begin{subfigure}[h]{.475\linewidth}
        \includegraphics[width=\textwidth]{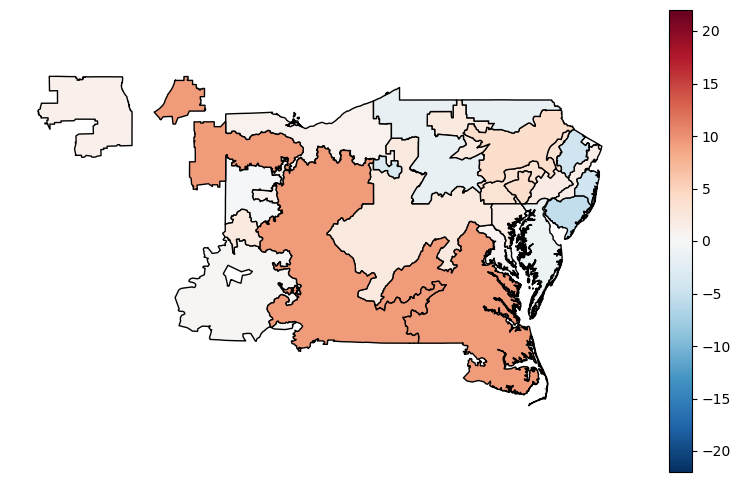}
        \caption{Co-optimized}
        \label{fig:net_load_coopt}
    \end{subfigure}
    \caption{Average Hourly Net Load 2045 [GW]}
    \label{fig:net_load}
\end{figure}

In Fig.\;\ref{fig:net_load}, we show the average net load in each zone for the last operational year. The net load is defined as the sum of zonal load and storage consumption minus zonal generation and unserved energy. Positive values indicate that the zone is, on average, importing energy and negative values that the zone is exporting energy. The co-optimized planning results depicted in Fig.\;\ref{fig:net_load_coopt}, show less extreme spatial average net load distribution compared to the sequentially planned system in Fig.\;\ref{fig:net_load_seq}. The result is consistent with the lower identified transmission need in the co-optimized planning model, as lower net load implies a reduced need for importing and exporting power.

\subsection{Impact on Costs, Reliability, and Emissions} \label{subsubsection:cost_rel_emis}
We find that in our main specification (`ELCC + High Gas Price + High Demand') the net present value system cost over the whole planning period is 3.7\% lower in the co-optimized planning model compared to the sequential planning model which amounts to approximately \$23 billion (2024\$) over the planning horizon (see Table \ref{tabular:cost_table_2}). Capacity investment rises slightly in the co-optimized scenario while operating costs fall, but the majority of the \$23 billion cost difference is due to a reduction in transmission costs. Since the transmission costs in the sequential case are \$26 billion and in the co-optimized case are \$4 billion, this amounts to a 85\% reduction in transmission costs. These transmission costs may be lower than those estimated by PJM both because they are expressed as net present values and because transmission expansion is assumed to occur through reconductoring which is a relatively low-cost way of expanding transmission. Unlike investments in generation and storage, the costs associated with transmission investments are typically solely borne by ratepayers. We see both total cost reductions and transmission cost reductions across sensitivities discussed in \ref{subsubsection:sensitivities}.

Our main scenario sees no difference in reliability outcomes (Table\;\ref{tabular:lost_load}). Across our sensitivities, co-optimization has no impact on reliability or slightly increase reliability. Our main scenario has 7\% decrease in emissions when moving from sequential to co-optimized planning, but across our sensitivities, co-optimized scenarios sometimes increase and sometimes decrease emissions while overall changes remain small (Table\;\ref{tabular:emis}).  

\subsection{Sensitivity Analyses} \label{subsubsection:sensitivities}
In this section, we present results for different modeling specifications. Specifically, we use a constraint that ensures reliability by requiring the model to generate enough to meet 115\% of demand in every hour instead of relying on an ELCC-based reliability constraint that requires enough ELCC rated capacity to meet 115\% of peak demand. We denote these alternate reliability constraint cases, `Rsv Req' because these cases enforce a reserve requirement in every hour. And we rerun scenarios with all combinations of low and high demand and gas prices.

Across sensitivities, the co-optimized planning model consistently builds less transmission capacity than the sequential model (see Table~\ref{tabular:trans_table}). This is because, as discussed in Section~\ref{subsubsection:trans_cap_res}, the co-optimization pushes the model to build capacity in more zones, closer to the load. Furthermore, ELCC-based reliability constraints reduce the transmission expansion need. ELCC ratings require gas to meet reliability but state policies require the building of renewables, so `ELCC' cases have roughly the same amount of renewable capacity as `Rsv Req' cases but much more total capacity. Since generation and transmission can serve as substitutes for one another, this larger total amount of generation reduces the need for transmission. \\

\begin{table} [h]
    \caption{Cumulative Transmission Build by 2045 [GW-mi] }    
    \label{tabular:trans_table}
    \begin{center}
    \begin{tabular}{ |l|c|c| } 
         \hline
         Scenarios                           & Sequential & Co-optimized \\ 
         \hline
         ELCC + High Gas Price + High Demand & 27,489 & 5,989 \\ 
         ELCC + High Gas Price + Low Demand  & 15,776 & 3,104 \\ 
         ELCC + Low Gas Price + High Demand  & 21,478 & 3,204 \\ 
         ELCC + Low Gas Price + Low Demand   & 7,766  & 563 \\
         \hline
         Rsv Req + High Gas Price + High Demand & 30,848 & 8,010 \\ 
         Rsv Req + High Gas Price + Low Demand  & 21,471 & 3,374 \\ 
         Rsv Req + Low Gas Price + High Demand  & 25,479 & 4,252 \\ 
         Rsv Req + Low Gas Price + Low Demand   & 10,780 & 2,271 \\         
         \hline
    \end{tabular}
    \end{center}
\end{table}

\begin{figure}
    \begin{subfigure}{.475\linewidth}
      \includegraphics[width=\linewidth]{figures/trans_seq0315.png}
      \caption{Sequential High Demand}
      \label{trans_seq_highDem}
    \end{subfigure}
    \begin{subfigure}{.475\linewidth}
      \includegraphics[width=\linewidth]{figures/trans_coopt0315.png}
      \caption{Co-optimized High Demand}
      \label{trans_coopt_highDem}
    \end{subfigure}
    \begin{subfigure}{.475\linewidth}
      \includegraphics[width=\linewidth]{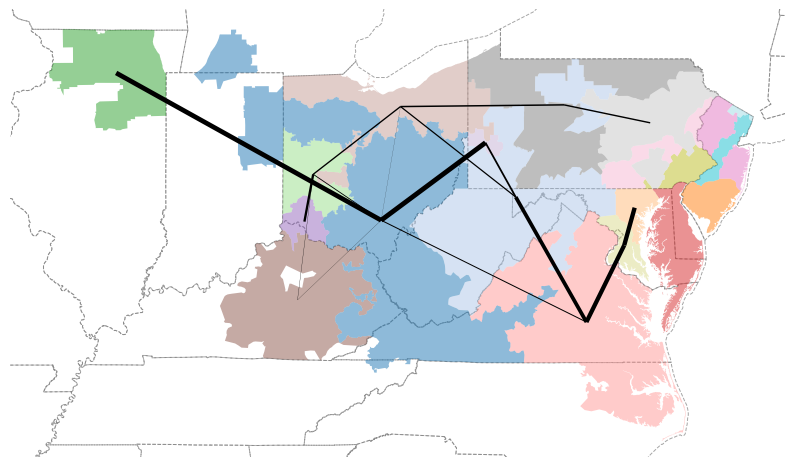}
      \caption{Sequential Low Demand}
      \label{trans_seq_lowDem}
    \end{subfigure}
    \begin{subfigure}{.475\linewidth}
      \includegraphics[width=\linewidth]{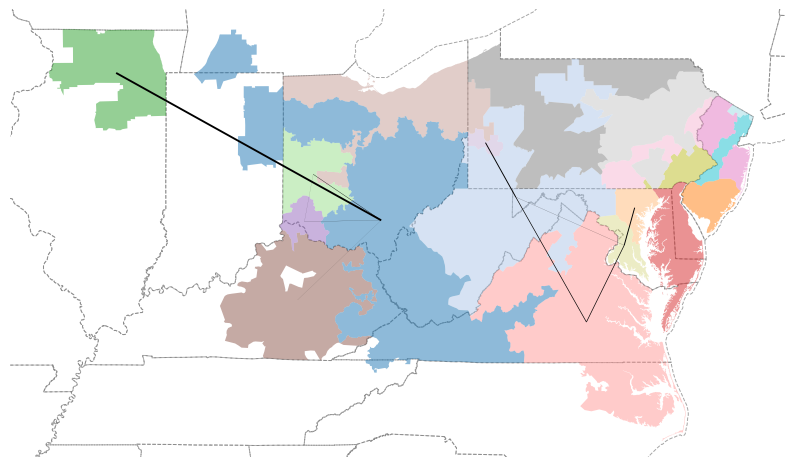}
      \caption{Co-optimized Low Demand}
      \label{trans_coopt_lowDem}
    \end{subfigure}
    \caption{Transmission Need Across Demand Scenarios}    
    \label{fig:load_trans_map}
\end{figure}

Transmission capacity is sensitive to gas prices and total demand, but co-optimization reduces transmission need regardless of the sensitivity. Higher demand typically requires more transmission than low demand cases. Higher gas prices also require more transmission, because those cases favor renewable capacity, which may be located farther from load. Under both circumstances, however, co-optimized planning reduces the need for transmission. Fig. \ref{fig:load_trans_map} shows high and low demand cases under sequential and co-optimized planning with high gas prices. While the high demand case requires more transmission than the low demand case, the difference in transmission between sequential and co-optimized cases is larger than between low and high demand cases.

We present the net present value of system cost for the co-optimized and the sequential transmission model using alternative scenarios in Table~\ref{tabular:cost_table_2}. The cost savings of the co-optimized scenarios in these sensitivities vary widely (`ELCC + Low Gas Price + Low Demand': 1.1\% ;`Rsv Req + Low Gas Price + High Demand': 30.7\%).

Reliability can be measured as total unserved energy, the cost of unserved energy assuming the value of lost load, or as unserved energy as a share of load. There is no reliability benefit in the `ELCC' cases because the `ELCC' cases require so much gas generation to be built that they have no lost load. But in the `Rsv Req' cases, the co-optimized scenario improves reliability (Table\;\ref{tabular:lost_load}).   

The co-optimized scenarios have similar emissions to the sequential scenarios (Table\;\ref{tabular:emis}). Six out of eight pairs of sensitivities have lower emissions under the co-optimized case, but two have higher emissions. Since co-optimization mostly allows a more cost effective use of generation and transmission resources, co-optimization is only likely to reduce emissions in scenarios where reducing emissions is cheaper, which depends on the relative cost of gas and renewables.

\begin{table} [h]
    \begin{center}
    \caption{NPV Cost 2025-2045 [Billion 2024\$] }    
    \label{tabular:cost_table_2}
    \begin{tabular} {|p{0.3\linewidth} | p{0.2\linewidth} | p{0.15\linewidth} |p{0.15\linewidth}|}
    \hline
    Scenarios                                  & Cost Metric  & Sequential & Co-optimized  \\
    \hline
    \hline

         \multirow{5}{25mm}{ELCC + High Gas Price + High Demand}
                                               & Operation    & 385 & 382 \\
                                               & UE           & 0   & 0   \\
                                               & Investment   & 208 & 211 \\
                                               & Transmission & 26  & 4   \\\cline{2-4}
                                               & System       & 620 & 597 \\                                               
         \hline                                             
         \multirow{5}{25mm}{ELCC + High Gas Price + Low Demand}
                                               & Operation    & 297 & 295 \\
                                               & UE           & 0   & 0   \\
                                               & Investment   & 139 & 141 \\         
                                               & Transmission & 14  & 3   \\\cline{2-4}
                                              & System       & 451 & 438 \\                                               
         \hline                                             
         \multirow{5}{25mm}{ELCC + Low Gas Price + High Demand}
                                               & Operation    & 327 & 325 \\
                                               & UE           & 0   & 0   \\
                                               & Investment   & 165 & 167 \\        
                                               & Transmission & 13  & 1   \\\cline{2-4}
                                               & System       & 505 & 494 \\                                               
         \hline                                             
         \multirow{5}{25mm}{ELCC + Low Gas Price + Low Demand}
                                               & Operation    & 260 & 260 \\
                                               & UE           & 0   & 0   \\
                                               & Investment   & 109 & 110 \\
                                               & Transmission & 5   & 0   \\\cline{2-4}
                                               & System       & 375 & 371 \\                                               
         \hline
         \hline
         \multirow{5}{25mm}{Rsv Req + High Gas Price + High Demand}
                                               & Operation    & 419 & 404 \\
                                               & UE           & 116 & 2   \\
                                               & Investment   & 204 & 204 \\         
                                               & Transmission & 30  & 6   \\\cline{2-4} 
                                               & System       & 768 & 616 \\                                               
         \hline                                             
         \multirow{5}{25mm}{Rsv Req + High Gas Price + Low Demand}
                                               & Operation    & 303 & 303 \\
                                               & UE           & 11  & 1   \\
                                               & Investment   & 153 & 154 \\         
                                               & Transmission & 20  & 3   \\\cline{2-4}
                                               & System       & 487 & 461 \\                                               
         \hline                                             
         \multirow{5}{25mm}{Rsv Req + Low Gas Price + High Demand}
                                               & Operation    & 334 & 325 \\
                                               & UE           & 180 & 1   \\
                                               & Investment   & 141 & 140 \\         
                                               & Transmission & 22  & 3   \\\cline{2-4}
                                               & System       & 677 & 469 \\                                               
         \hline                                             
         \multirow{5}{25mm}{Rsv Req + Low Gas Price + Low Demand} 
                                               & Operation    & 253 & 254 \\
                                               & UE           & 8   & 0   \\
                                               & Investment   & 104 & 104 \\	        
                                               & Transmission & 11  & 2   \\\cline{2-4}
                                               & System       & 376 & 360 \\                                               
         \hline                                             
    \hline
    \end{tabular}
    \end{center}
\end{table}

\begin{table} [h]
    \caption{Reliability Metrics}    
    \label{tabular:lost_load}
    \begin{center}
    \begin{tabular} {|p{0.3\linewidth} | p{0.2\linewidth} | p{0.15\linewidth} |p{0.15\linewidth}|}
    \hline
    Scenarios                             & Metric         & Sequential & Co-optimized  \\
    \hline
    \hline
    \multirow{3}{28mm}{ELCC + High Gas Price + High Demand}
                                          & UE [GWh]       & 0.     & 0.  \\  
                                          & NPV UE [B\$]   & 0.     & 0.  \\ 
                                          & UE/Load [\%]   & 0.     & 0. \\ 
    \hline
    \multirow{3}{28mm}{ELCC + High Gas Price + Low Demand}
                                         & UE [GWh]       & 0.     & 0.     \\ 
                                         & NPV UE [B\$]   & 0.     & 0.     \\ 
                                         & UE/Load [\%]   & 0.     & 0.     \\ 
    \hline
    \multirow{3}{28mm}{ELCC + Low Gas Price + High Demand}    
                                         & UE [GWh]       & 0.     & 0.     \\ 
                                         & NPV UE [B\$]   & 0.     & 0.     \\ 
                                         & UE/Load [\%]   & 0.     & 0.     \\  
    \hline
    \multirow{3}{28mm}{ELCC + Low Gas Price + Low Demand}    
                                         & UE [GWh]       & 0.     & 0.     \\  
                                         & NPV UE [B\$]   & 0.     & 0.     \\ 
                                         & UE/Load [\%]   & 0.     & 0.     \\ 
    \hline
    \hline
    \multirow{3}{28mm}{Rsv Req + High Gas Price + High Demand}     
                                          & UE [GWh]       & 42,655  & 1,093  \\ 
                                          & NPV UE [B\$]   & 115.875 & 2.319 \\ 
                                          & UE/Load [\%]   & 0.184   & 0.005 \\ 
    \hline
    \multirow{3}{28mm}{Rsv Req + High Gas Price + Low Demand}    
                                         & UE [GWh]       & 4,054     & 387    \\ 
                                         & NPV UE [B\$]   & 10.85    & 0.815  \\ 
                                         & UE/Load [\%]   & 0.023    & 0.002  \\ 
    \hline
    \multirow{3}{28mm}{Rsv Req + Low Gas Price + High Demand}    
                                         & UE [GWh]       & 86,434     & 654   \\ 
                                         & NPV UE [B\$]   & 179.517    & 1.31  \\ 
                                         & UE/Load [\%]   & 0.373      & 0.003 \\  
    \hline
    \multirow{3}{28mm}{Rsv Req + Low Gas Price + Low Demand}    
                                         & UE [GWh]       & 3,217     & 253     \\  
                                         & NPV UE [B\$]   & 7.874    &  0.471  \\ 
                                         & UE/Load [\%]   & 0.018       & 0.001\\ 
    \hline
    \end{tabular}
    \end{center}
\end{table}

\begin{table} [h]
    \caption{Emissions 2025-2045 [million metric tons] }    
    \label{tabular:emis}
    \begin{center}
    \begin{tabular}{ |l|c|c| } 
         \hline
         Scenarios                           & Sequential & Co-optimized \\ 
         \hline
         ELCC + High Gas Price + High Demand & 3,154   & 2,903   \\ 
         ELCC + High Gas Price + Low Demand  & 2,259   & 2,196   \\ 
         ELCC + Low Gas Price + High Demand  & 4,400   & 4,241   \\ 
         ELCC + Low Gas Price + Low Demand   & 2,802   & 2,769   \\ 
         \hline
         Rsv Req + High Gas Price + High Demand & 4,273   & 4,149    \\ 
         Rsv Req + High Gas Price + Low Demand  & 2,706   & 2,727   \\  
         Rsv Req + Low Gas Price + High Demand  & 5,915   & 5,659   \\  
         Rsv Req + Low Gas Price + Low Demand   & 3,813   & 3,816   \\         
         \hline
    \end{tabular}
    \end{center}
\end{table}

\section{Conclusion} \label{sec:conclusion}
We compare transmission planning outcomes using a sequential planning model for the PJM region with a coordinated planning model in which generation, storage, and transmission is co-optimized. We find that the latter leads to significantly less transmission need and lower total system costs with similar reliability performance and carbon dioxide emissions. 

The co-optimized planning approach does not eliminate the need for detailed power flow studies based on a realistic representation of the grid. We expect that this will create additional transmission upgrade requirements beyond the corridor upgrades we have identified. However, our simplified approach provides a strategic framework for identifying and expanding transmission corridors, which can then be further refined into specific transmission elements through more detailed engineering analyses.

Future research could focus on stochastic models to accurately capture the operational uncertainties but also deep uncertainties related to demand growth and technology innovation.

\bibliographystyle{IEEEtran}
{\footnotesize \bibliography{IEEEabrv,main}}

\begin{appendix} \label{appendix:ap}
    \section{Appendix}\label{sec:model}

This appendix describes the model. It begins with nomenclature, then gives the model formulation for co-optimized grid planning and the variations that permit sequential grid planning. It also includes a description of the model constrains used for a capacity market style reliability requirement. 

\subsection{Nomenclature}
\newcommand{\longestitem}{$instate\_rps^{all-re}_{e,st}$}
\subsubsection{Indices and Sets}
\begin{ldescription}{\longestitem}
\item [$e \in E$] epochs
\item [$e_i$] initial year of an epoch i.e. year in which investment decisions are made
\item [$t \in T$] years from start of first epoch ($e_i$ of $e=0$) to end of last epoch ($e_f$ of $e=E$) 
\item [$g \in G$] generation units
\item [$s \in S$] storage units
\item [$h \in H$] hours
\item [$l \in L$] transmission corridors
\item [$z \in Z$] zones 
\item [$st \in ST$] states 
\item [$st^{g} \in ST^{g}$] states with generator capacity targets where $ST^{g} \subseteq ST$ 
\item [$sts^{g} \in STS^{g}$] where $STS^{g}$ is the power set (set of all subsets $sts^{g}$) of $ST^{g}$ ($STS^{g} = P(ST^{g})$)
\item [$st^{s} \in ST^{s}$] states with storage capacity targets where $ST^{s} \subseteq ST$ 
\item [$sts^{s} \in STS^{s}$] where $STS^{s}$ is the power set (set of all subsets $sts^{s}$) of $ST^{s}$ ($STS^{s} = P(ST^{s})$)
\item [$p \in P$] where each $p$ is a subset of $g$ eligible for compliance with an RPS policy. In this modeling effort the subsets of $g$ with corresponding RPS policies are $g^{all-re}$,$g^{re}$,$g^{pv}$, and $g^{wind}$
\item [$g^{all-re}$] all renewable generation units (solar, offshore wind, onshore wind, hydro)
\item [$g^{re} \in G$] renewable generation units (solar, offshore wind, onshore wind) 
\item [$g^{pv} \in G$] solar generating units
\item [$g^{wind}\in G$] wind generating units including both offshore and onshore wind
\end{ldescription}

\subsubsection{Parameters}
\begin{ldescription}{\longestitem}
\item [$d$] duration of each epoch
\item [$r$] discount rate with respect to time
\item[$\delta\_wgt_{e}$] discounted weight applied to operating costs of each epoch
\item[$var\_cost^{gen}_{e,z,g}$] variable cost of each generator ($g$) in epoch ($e$), zone ($z$)
\item[$fom^{gen}_{e,z,g}$] fixed O\&M cost of each generator ($g$) in epoch ($e$), zone ($z$)
\item[$inv\_cost^{gen}_{e,z,g}$] disounted capital costs of generators
\item[$cap^{gen}_{e,z,g}$] initial capacity of generators at the start of an epoch (an exogenously fixed quantity)
\item[$cap\_fact^{gen}_{e,z,g,h}$] hourly capacity factor for generators
\item[$load_{e,z,h}$] load in each zone, epoch, and hour
\item[$VoLL$] value of lost load [$\$/MWh$]
\item[$inv\_cost^{stor}_{e,z,s}$] disounted capital costs of storage
\item[$cap^{stor}_{e,z,s}$] initial (power) capacity of storage at the start of an epoch
\item[$energy\_cap^{stor}_{e,z,s}$] initial energy capacity of storage at the start of an epoch
\item[$dur^{stor}_{s}$] the factor used to convert storage power into storage energy aka storage duration
\item[$\eta_{s}$] round trip efficiency of storage units
\item[$inv\_cost^{line}_{e,l}$] disounted capital costs of transmission
\item[$cap\_lb^{line}_{e,l}$] initial lower bound on transmission capacity across a given line
\item[$cap\_ub^{line}_{e,l}$] initial upper bound on transmission capacity across a given line
\item[$map^{line}_{e,z,l}$] mapping between transmission corridors and the zones they connect
\item[$reinf\_fact^{line}$] maximum reinforcement factor for transmission corridors
\item[$instate\_rps^{all-re}_{e,st}$] percent of an rps that must be met within-state
\item[$state\_rps^{all-re}_{e,st}$] rps percentage
\item[$zone\_share_{st,z}$] share of each zone belonging to each state
\item[$region\_rps_{e,p}$] regional RPS requirement for subset of renewable generators 
\item[$map^{state}_{st,z}$] map states to zones they overlap (values 1,0)
\item[$req^{ofs}_{e,st}$] offshore wind capacity requirement by state
\item[$req^{stor}_{e,st}$] storage capacity requirement by state
\item[$req^{pv}_{e,st}$] solar capacity requirement by state
\item[$rsv$] reserve factor, typically 15\%
\item[$elcc^{gen}_{e,g}$] Projected ELCC for generators
\item[$elcc^{stor}_{e,s}$] Projected ELCC for storage units
\item[$cap\_target_{e}$] target of the capacity market
\item[$limit^{gen}_{z,g}$] cumulative limit on generation capacity for each zone. Only implemented for solar, gas-cc, onshore wind
\item[$limit^{stor}_{zs}$] cumulative limit on storage capacity for each zone\\
\end{ldescription}

\subsubsection{Positive Continuous Variables}
\begin{ldescription}{\longestitem}
\item [$\Phi$] net present value of system cost
\item [$X^{gen}_{e,z,g}$] capacity of generator ($g$) added in epoch ($e$), zone ($z$)
\item [$X^{gen\_retire}_{e,z,g}$] capacity of generator ($g$) exogenously retired in epoch ($e$), zone ($z$)
\item [$X^{stor}_{e,z,s}$] (power) capacity of storage ($s$) added in epoch ($e$), zone ($z$)
\item [$X^{line}_{e,l}$] transmission capacity added on line ($l$) in epoch ($e$) 
\item [$Q^{gen}_{e,z,g,h}$] dispatch of generation unit ($g$) in each epoch ($e$), zone ($z$), hour ($h$)
\item [$Q^{stor}_{e,z,s,h}$] dispatch of storage unit ($s$) in each epoch ($e$), zone ($z$), hour ($h$)
\item [$CONS^{stor}_{e,z,s,h}$] consumption of storage unit ($s$) in each epoch ($e$), zone ($z$), hour ($h$)
\item [$SOC^{stor}_{e,z,s,h}$] sate-of-charge of storage unit ($s$) in each epoch ($e$), zone ($z$), hour ($h$)
\item [$UE_{e,z,h}$] unserved energy in each epoch ($e$), zone ($z$), hour ($h$)
\item [$Q\_REC_{e,z,g^{all-re}}$] RECs generated in epoch ($e$), zone ($z$), by renewable generator $g^{all-re}$ \\
\end{ldescription}

\subsubsection{Continuous Variables}
\begin{ldescription}{\longestitem}
\item [$FLOW_{e,l,h}$] power flow on transmission corridor ($l$) in epoch ($e$) and hour ($h$)
\end{ldescription}

\subsection{Model Description}
The model solves for generation-, storage-, and transmission-capacity additions, retirements of existing generation, as well as system operations over a 20-year time horizon. The time horizon is divided into four five-year epochs ($e$). In the first year of each epoch, the model makes investment decisions, and in the final year of each epoch it makes operations decisions for 8760 hours including any new capacity from the investments.

The model represents the electric system as a set of zones ($z$) connected by transmission corridors ($l$). Each zone has legacy generation and storage technologies ($g$ and $s$) with some existing capacity described by ($cap^{gen}_{e,z,g}$ and $cap^{stor}_{e,z,s}$) and any increase in that capacity chosen in the capacity variables ($X^{gen}_{e,z,g}$ and $X^{stor}_{e,z,s}$). Retirements of legacy generation are modeled endogenously ($X^{gen\_retire}_{e,z,g}$). The transmission corridor capacities have an existing capacity ($cap\_lb^{line}_{e,l}$ and $cap\_ub^{line}_{e,l}$), which can be expanded by investing in new capacity ($X^{line}_{e,l}$). Operating decisions are tracked for generators with the generation variables ($Q^{gen}_{e,z,g,h}$); for storage with the storage variables for generation ($Q^{stor}_{e,z,s,h}$), consumption ($CONS^{stor}_{e,z,s,h}$), and state-of-charge ($SOC^{stor}_{e,z,s,h}$); and for transmission with a variable to record hourly flow along corridors ($FLOW_{e,l,h}$). Although the model has a parametric load for each zone ($load_{e,z,h}$), the model can choose not to meet that load if it pays a penalty equivalent to the value of lost load ($VoLL$) of \$5000 (2024\$) for unserved energy. The quantity of unserved energy is stored in the decision variable ($UE_{e,z,h}$).

\subsection{Co-optimized Planning Model} \label{subsection:coopt_model}
\subsubsection{Objective Function}
The objective function (\ref{equation:obj_func}) minimizes the discounted sum of operating costs (the cost of generation and unserved energy) and investment costs (the cost of additional generation, storage, and transmission capacity). Operations in the final year of each epoch are weighted to stand-in for the five years of the epoch ($\delta\_wgt_{e}$). This weight is the discount factor appropriate to each year, but summed across the five years of the epoch (\ref{equation:delta_wgt}). For investment costs, we follow the discounted investment cost method described in \cite{LARA20181037}. We first calculate the annualized cost for each technology ($acc_{e}$), then calculate the time-discounted sum of annual costs to the end of the model's time horizon as in Equation (\ref{equation:dcc}).

\begin{equation}
    \label{equation:obj_func}
    \begin{aligned}
        min\;\Phi &=  \sum_{e,z,g,h}{Q^{gen}_{e,z,g,h} \times var\_cost^{gen}_{e,z,g} \times \delta\_wgt_{e}}\\
        &+ \sum_{e,z,h}{UE_{e,z,h} \times VoLL \times \delta\_wgt_{e}}\\
        &+ \sum_{e,z,g}{X^{gen}_{e,z,g} \times inv\_cost^{gen}_{e,z,g}}\\
        &+ \sum_{e,z,s}{X^{stor}_{e,z,s} \times inv\_cost^{stor}_{e,z,s}}\\
        &+ \sum_{e,l}{X^{line}_{e,l} \times inv\_cost^{line}_{e,l}}\\
        &+ \sum_{e,z,g}\Biggl( \biggl( cap^{gen}_{e,z,g} - X^{gen\_retire}_{e,z,g} + \sum^{e}_{e=0,z,g}{X^{gen}_{e,z,g}} \biggr)  \\
        &~~~~~~~~~\times fom^{gen}_{e,z,g} \times \delta\_wgt_{e} \Biggr)\\
    \end{aligned}
\end{equation}

\begin{equation}
    \label{equation:delta_wgt}
    \begin{aligned}
        \delta\_wgt_{e} = \sum_{t=e_i}^{t=e_{f}}{\frac{1}{(1+r)^t}}
    \end{aligned}
\end{equation} 

\begin{equation}
    \label{equation:dcc}
    \begin{aligned}
        inv\_cost_{e} = acc_{e} \times \sum_{t=e_i}^{t=T}\frac{1}{(1+r)^t}
    \end{aligned}
\end{equation} 

\subsubsection{Generation}
The model requires that the generation of each generator is less than its maximum available capacity in each hour, including the capacity growth (\ref{equation:gen_le_cap}). It also forbids the retirement of more capacity than exists (\ref{equation:ret_ex}). And it requires that once retired, capacity cannot come back into service  (\ref{equation:ret_ex_across_time}).
\begin{equation}
    \label{equation:gen_le_cap}
    \begin{aligned}
        (cap^{gen}_{e,z,g} - X^{gen\_retire}_{e,z,g} + \sum_{e=0}^{e}{X^{gen}_{e,z,g}}) &\times cap\_fact^{gen}_{e,z,g} \\
        &\geq Q^{gen}_{e,z,g,h}\\
        \forall_{e,z,g,h}
    \end{aligned}
\end{equation}

\begin{equation}
    \label{equation:ret_ex}
    \begin{aligned}
        cap^{gen}_{e,z,g} - X^{gen\_retire}_{e,z,g} \geq 0\\
        \forall_{e,z,g}
    \end{aligned}
\end{equation}

\begin{equation}
    \label{equation:ret_ex_across_time}
    \begin{aligned}
        (cap^{gen}_{e,z,g} - X^{gen\_retire}_{e,z,g}) \\
        - (cap^{gen}_{e+1,z,g} - X^{gen\_retire}_{e+1,z,g}) \\
        \geq 0\\
        \forall_{e<E,z,g}
    \end{aligned}
\end{equation}

\subsubsection{Energy Storage}
The model requires that the rate at which storage generates electricity must be less than the power capacity of the storage, including capacity growth (\ref{equation:gen_le_cap_stor}). It also requires that the rate at which storage consumes electricity must be less than the power capacity of the storage including capacity growth (\ref{equation:cons_le_cap_stor}). To guarantee intertemporal consistency and energy balance for storage, the model tracks the sate-of-charge ($SOC^{stor}_{e,z,s,h}$). Equation (\ref{equation:soc_le_cap_stor}) requires that the state-of-charge be less than the energy capacity of the storage including energy capacity growth. Equations \eqref{equation:soc_i} and \eqref{equation:soc_f} set the initial and final sate-of-charge over a year to both equal half the energy capacity to ensure energy conservation. Equation (\ref{equation:soc_m}) tracks the sate-of-charge of the battery in each hour so that each hour the charge depends on the sate-of-charge in the previous hour and the quantity of charging and discharging occurring including efficiency losses.

\begin{equation}
    \label{equation:gen_le_cap_stor}
    \begin{aligned}
            (cap^{stor}_{e,z,s} + \sum_{e=0}^{e}{X^{stor}_{e,z,s}}) \geq Q^{stor}_{e,z,s,h}\\
            \forall_{e,z,s,h}
    \end{aligned}
\end{equation}

\begin{equation}
    \label{equation:cons_le_cap_stor}
    \begin{aligned}
        (cap^{stor}_{e,z,s} + \sum_{e=0}^{e}{X^{stor}_{e,z,s}}) \geq CONS^{stor}_{e,z,s,h}\\
        \forall_{e,z,s,h}
    \end{aligned}
\end{equation}

\begin{equation}
    \label{equation:soc_le_cap_stor}
    \begin{aligned}
    (\sum_{e=0}^{e}{X^{stor}_{e,z,s}} \times dur^{stor}_{e,z,s}) + energy\_cap^{stor}_{e,z,s}\\
    \geq SOC^{stor}_{e,z,s,h}\\
    &&\forall_{e,z,s,h}
    \end{aligned}
\end{equation}

\begin{equation}
    \label{equation:soc_i}
    \begin{aligned}
        SOC^{stor}_{e,z,s,0} = & 0.5 \times (energy\_cap^{stor}_{e,z,s} \\
        & + (\sum_{e=0}^{e}{X^{stor}_{e,z,s}} \times dur^{stor}_{e,z,s}))\\
        & + CONS^{stor}_{e,z,s,0}- Q^{stor}_{e,z,s,0}\\
        &&\forall_{e,z,s}
    \end{aligned}
\end{equation}

\begin{equation}
    \label{equation:soc_f}
    \begin{aligned}
        SOC^{stor}_{e,z,s,H} = & 0.5 \times (energy\_cap^{stor}_{e,z,s} \\
        &+ (\sum_{e=0}^{e}{X^{stor}_{e,z,s}} \times dur^{stor}_{e,z,s}))\\
        &&\forall_{e,z,s}
    \end{aligned}
\end{equation}

\begin{equation}
    \label{equation:soc_m}
    \begin{aligned}
        SOC^{stor}_{e,z,s,h} = & SOC^{stor}_{e,z,s,h-1} \\
        &+ (CONS^{stor}_{e,z,s,h} \times \eta_{s})\\
        &- Q^{stor}_{e,z,s,h}\\
        &&\forall_{e,z,s,h>0}
    \end{aligned}
\end{equation}

\subsubsection{Transmission Constraints}
The basic version of the model includes a transmission network with an initial capacity and the option to expand the capacity of that network. Equations (\ref{equation:flow_le_cap_ub}) and (\ref{equation:flow_le_cap_lb}) ensure that the power flowing across a corridor in any given hour is less than the capacity of that line including capacity growth. Power flow is modeled as a decision variable without enforcing Kirchhoff's laws. 

\begin{equation}
    \label{equation:flow_le_cap_ub}
    \begin{aligned}
        (cap\_ub^{line}_{e,l} + \sum_{e=0}^{e}{X^{line}_{e,l}}) \geq FLOW_{e,l,h}\\
        &&\forall_{e,l,h}
    \end{aligned}
\end{equation}

\begin{equation}
    \label{equation:flow_le_cap_lb}
    \begin{aligned}
        FLOW_{e,l,h} \geq -(cap\_lb^{line}_{e,l} + \sum_{e=0}^{e}{X^{line}_{e,l}})\\
        &&\forall_{e,l,h}
    \end{aligned}
\end{equation}




\subsubsection{Power Balance} \label{subsubsection:power_balance}

The power balance constraint (\ref{equation:sup_dem_eq}) requires that in each zone and in each hour, there is enough electricity to meet load plus the reserve margin by requiring the sum of generation, storage generation, unserved energy, and net imports to a location to equal load at each hour and location. Equation (\ref{equation:ue_lim}) limits unserved energy to be no greater than load. 

\begin{equation}
    \label{equation:sup_dem_eq}
    \begin{aligned}
        load_{e,z,h} * (1 + rsv) = &\sum_{g}{Q^{gen}_{e,z,g,h}}\\
        &+ \sum_{s}{Q^{stor}_{e,z,s,h}}\\
        &- \sum_{s}{CONS^{stor}_{e,z,s,h}}\\
        &+ UE_{e,z,h}\\
        &+ \sum_{l}(map^{line}_{e,z,l} \times FLOW_{e,l,h})\\
        &&\forall_{e,z,h}
    \end{aligned}
\end{equation}

\begin{equation}
    \label{equation:ue_lim}
    \begin{aligned}
        load_{e,z,h} - UE_{e,z,h} \geq 0\\
        &&\forall_{e,z,h}
    \end{aligned}
\end{equation}

\subsubsection{Capacity Limits} \label{subsubsection:cap_lim}
Equations \eqref{equation:queue_lim} and \eqref{equation:queue2_lim} limit cumulative capacity growth by zone.

\begin{equation}
    \label{equation:queue_lim}
    \begin{aligned}
        limit^{gen}_{z,g} \geq \sum_{e=0}^{e=E}{X^{gen}_{e,z,g}}\\
        &&\forall_{z,g}
    \end{aligned}
\end{equation}
  
\begin{equation}\label{equation:queue2_lim}
    \begin{aligned}
        limit^{stor}_{zs} \geq \sum_{e=0}^{e=E}{X^{stor}_{e,z,s}}\\
        &&\forall_{z,s}
    \end{aligned}
\end{equation}

\subsubsection{Policy Constraints} \label{subsubsection:pol_const}
The model also has constraints to represent state policies such as Renewable Portfolio Standards (RPS) and renewable and storage capacity targets. To enable the enforcement of state RPS policies, the variable ($Q\_REC_{e,z,g}$) tracks renewable generation. Equation (\ref{equation:rec_def}) defines the number of renewable energy credits of a given renewable generation type generated annually in a zone to be equal to the amount of generation of that type in the zone. Several of the PJM states have RPS, some of which must be met with renewable generation within the state and some of which can be met with renewable generation anywhere in the region. Equation \eqref{equation:rps_all} is the region-wide RPS constraint which requires enough renewable generation of the specified variety to occur in the region to meet state policies. Equation \eqref{equation:rps_inst} is the in-state RPS requirement (primarily for Virginia and Illinois which require a share of their state RPS to be met with in-state generation). This constraint assigns a portion of the state's RPS to each zone that overlaps the state proportional to the zone's share of the state's load. Thus there is one version of the constraint for each zone.

State capacity target constraints are given by Equation (\ref{equation:gen_req}) for offshore wind and solar generation and (\ref{equation:stor_req}) battery storage. To avoid double counting of capacity that appears in zones that overlap multiple states with capacity targets for the same technology, equations (\ref{equation:gen_req}) and (\ref{equation:stor_req}) are indexed over $sts^{g}\in STS^{g}$ and $sts^{s}\in STS^{s}$ where $sts$ is a set of states that all have capacity target policies for a given technology and $STS$ is the set of all combinations of states that have capacity target policies for a given technology. In this way, we require not only that there be enough capacity to meet a state target in all zones overlapping a state, but we also require that for any group of states with targets, there must be enough capacity in all zones overlapping those states to meet the sum of those state targets.  

\begin{equation}
    \label{equation:rec_def}
    \begin{aligned}
        Q\_REC_{e,z,g^{all-re}} = \sum_{h}Q^{gen}_{e,z,g^{all-re},h}\\
        &&\forall_{e,z,g^{all-re}}
    \end{aligned}
\end{equation}


\begin{equation}
    \label{equation:rps_all}
    \begin{aligned}
        \sum_{z,g\in p}{Q\_REC_{e,z,g}} \geq region\_rps_{e,p}\\
        &&\forall_{e,p}
    \end{aligned}
\end{equation}

\begin{equation}
    \label{equation:rps_inst}
    \begin{aligned}
        \sum_{g^{all-re}}{Q\_REC_{e,z,g^{all-re}}} \geq& \sum_{h}{load_{e,z,h}}\\
        & \times \sum_{st}(instate\_rps^{all-re}_{e,st} \\
        & \times state\_rps^{all-re}_{e,st} \\
        & \times zone\_share_{st,z})  \\
        &&\forall_{e,z}
    \end{aligned}
\end{equation}


\begin{equation}
    \label{equation:gen_req}
    \begin{aligned}
        \sum_{z,g}{\biggl((cap^{gen}_{e,z,g} + \sum_{e=0}^{e}{X^{gen}_{e,z,g}}) \times \min(1,\sum_{st^{g} \in sts^{g}}map^{state}_{st^{g},z})} \biggr) \\
        \geq \sum_{st^{g} \in sts^{g}}{req^{g}_{e,st^{g}}}\\
        \forall_{e\in E,sts^{g}\in STS^{g}}
    \end{aligned}
\end{equation} 

\begin{equation}
    \label{equation:stor_req}
    \begin{aligned}
        \sum_{z,s}{\biggl( (cap^{stor}_{e,z,s} + \sum_{e=0}^{e}{X^{stor}_{e,z,s}})\times \min(1,\sum_{st^{s} \in sts^{s}}map^{state}_{st^{s},z})} \biggr) \\
        \geq \sum_{st^{s} \in sts^{s}}{req^{stor}_{e,st^{s}}}\\
        \forall_{e\in E,sts^{s}\in STS^{s}}
    \end{aligned}
\end{equation}

\subsection{Sequential Planning Model} \label{subsection:copper_plate}

For the sequential version of the model, we first solve the model in a `copper plate' form that excludes transmission considerations from the model, then re-solve the model fixing generation and storage capacity to `copper plate' levels and allowing transmission expansion. The `copper plate' generation and storage expansion planning model (GSEP) replaces the original objective function (\ref{equation:obj_func}) with a new objective function (\ref{equation:obj_func_copper_plate}) without transmission corridor capacity variables. It drops all equations related to transmission, i.e., \eqref{equation:flow_le_cap_ub} and \eqref{equation:flow_le_cap_lb}. And it replaces the zonal load balance constraint (\ref{equation:sup_dem_eq}) with a system-wide load-balance constraint (\ref{equation:sup_dem_reg}).

We then re-run the original version of the model with the difference that the generation and storage capacity decision variables are fixed to the corresponding generation and storage investment capacity decisions from the GSEP model. Under the fixed generation and storage capacity assumption, the model is effectively a reactive transmission expansion planning model (TEP).

\begin{equation}
    \label{equation:obj_func_copper_plate}
    \begin{aligned}
        obj\_func &=  \sum_{e,z,g,h}{Q^{gen}_{e,z,g,h} \times var\_cost^{gen}_{e,z,g} \times \delta\_wgt_{e}}\\
        &+ \sum_{e,z,h}{UE_{e,z,h} \times VoLL \times \delta\_wgt_{e}}\\
        &+ \sum_{e,z,g}{X^{gen}_{e,z,g} \times inv\_cost^{gen}_{e,z,g}}\\
        &+ \sum_{e,z,s}{X^{stor}_{e,z,s} \times inv\_cost^{stor}_{e,z,s}}\\
        &+ \sum_{e,z,g}\Biggl( \biggl( cap^{gen}_{e,z,g} - X^{gen\_retire}_{e,z,g} + \sum^{e}_{e=0,z,g}{X^{gen}_{e,z,g}}\biggr) \\
        & ~~~~~~~~~\times fom^{gen}_{e,z,g} \times \delta\_wgt_{e} \Biggr)\\
    \end{aligned}
\end{equation}

\begin{equation}
    \label{equation:sup_dem_reg}
    \begin{aligned}
        \sum_{z}{load_{e,z,h} * (1 + rsv)} = &\sum_{z,g}{Q^{gen}_{e,z,g,h}}\\
        &+ \sum_{z,s}{Q^{stor}_{e,z,s,h}}\\
        &- \sum_{z,s}{CONS^{stor}_{e,z,s,h}}\\
        &+ \sum_{z}{UE_{e,z,h}}\\
        &&\forall_{e,h}
    \end{aligned}
\end{equation}

\subsection{Other Model Options: Reliability Constraints} \label{subsection:reliability}
To check the robustness of our results we run the model with and without reliability constraints that mirror PJM's capacity market. This reliability constraint (\ref{equation:cap_market}) requires the region to have a certain target level of capacity ($cap\_target$), here defined at 115\% of peak load, where each resource has capacity accreditation based on PJM's published ELCCs. This version of the model also sets the reserve margin $rsv=0$ in (\ref{equation:sup_dem_eq}) (or (\ref{equation:sup_dem_reg}) in the `copper plate' model) so that the equation requires exactly enough generation to meet demand and thus mimics the behavior of the wholesale market. 

\begin{equation}
    \label{equation:cap_market}
    \begin{aligned}
        \sum_{z,g}{\biggl( (cap^{gen}_{e,z,g} - X^{gen\_retire}_{e,z,g} + \sum_{e=0}^{e}{X^{gen}_{e,z,g}}) * elcc^{gen}_{e,g}\biggr)}&\\
        + \sum_{z,s}{\biggr((cap^{stor}_{e,z,s} + \sum_{e=0}^{e}{X^{stor}_{e,z,s}}) * elcc^{stor}_{e,s}\biggr)}&\\
        \geq cap\_target_{e}&\\
        \forall_{e}\\
    \end{aligned}
\end{equation}

\end{appendix}

\end{document}